\documentclass[a4paper]{jpconf}
\usepackage{graphicx}

\begin{document}

\title{The phase diagram of $N_f=2$ and $N_f=2+1$ QCD from quark and gluon propagators}

\author{Jan Luecker and Christian S. Fischer}

\address{Institut f\"ur Theoretische Physik,
  Justus-Liebig-Universit\"at Gie\ss{}en,
  Heinrich-Buff-Ring 16,
  D-35392 Gie\ss{}en, Germany}

\begin{abstract}
In this talk we present results for the propagators and the phase diagram of QCD,
obtained from a truncated set of Dyson-Schwinger equations at finite temperature $T$
and quark chemical potential $\mu$.
We include back-coupling effects of the dressed quark and gluon propagators,
which also allows us to study the influence of strange quarks.
For the phase diagram we find a critical end-point at $\mu/T \approx 1.9$
and coinciding chiral and deconfinement phase transitions.
\end{abstract}

\section{Introduction}
The most reliable source of information on quantum chromodynamics (QCD) at finite
temperature has been lattice QCD which finds a first order phase transition in
the quenched case and a crossover when dynamical quarks are included \cite{Borsanyi:2010bp,Bazavov:2011nk}.
In the latter case the order parameter for confinement, the Polyakov loop,
shows a crossover transition nearby the chiral phase transition.
However, at finite density lattice studies are hindered by the fermion sign problem.
This led to a large industry of studies in effective models such as the Polyakov-loop extended
Nambu--Jona-Lasinio \cite{Fukushima:2003fw,Ratti:2005jh} and quark-meson models
\cite{Schaefer:2007pw,Skokov:2010wb,Herbst:2010rf}.
From these studies the general expectation that
the crossover turns into a first order phase transition at a critical end-point (CEP)
has emerged, with the position and even the existence of the CEP under debate.
When employing effective models one has to trade the QCD degrees of freedom
for effective ones.
In contrast to that we use Dyson-Schwinger equations (DSEs) to study
QCD and its degrees of freedom, quarks and gluons, directly.
We take advantage of this possibility by introducing a truncation where the back-reaction
of quarks onto gluons is explicitly taken into account.
Here we summarize our results published in Ref.~\cite{Fischer:2012vc}.

\section{Order parameters from Dyson-Schwinger equations}
The main objects of interest in this work are the propagators for the fully dressed quark
and gluon, which are given, respectively, as
\begin{eqnarray}
S(p) &=& [i(\omega_n+i\mu)\gamma_4C(p)+i\vec{p}\vec{\gamma}A(p)+B(p)]^{-1}\,,\nonumber\\ \label{eq:qProp}\\
D_{\mu\nu}(p) &=& P_{\mu\nu}^L(p)\frac{Z^L(p)}{p^2} + P_{\mu\nu}^T(p)\frac{Z^T(p)}{p^2}\,,
\end{eqnarray}
where $p=(\vec{p},\omega_p)$.
The quark is dressed by the scalar functions $A$, $B$ and $C$ which depend on $\vec{p}^2$
and $\omega_p$ only. The in-medium gluon has two dressing functions $Z_L$ and $Z_T$
for the components longitudinal and transversal to the medium.
$P_{\mu\nu}^L$ and $P_{\mu\nu}^T$ are the corresponding projectors.
In the quark propagator we neglected a fourth dressing functions which comes with
$\vec{p}\vec{\gamma}\gamma_4$ and can be shown by explicit calculations to have virtually no
influence on the results.

\subsection{Quark condensate}
In the quark propagator, Eq.~(\ref{eq:qProp}), the presence of chiral symmetry breaking
is directly reflected in a non-vanishing scalar dressing funtion $B$.
A derived order parameter for chiral symmetry breaking is the condensate

\begin{equation} \label{eq:condensate}
\langle\bar{\psi}\psi\rangle = 
Z_2T\sum_n\int\frac{d^3p}{(2\pi)^3}\mathrm{Tr}_D\left[S(p)\right],
\end{equation}
which is calculated from the propagator once the quark DSE is solved.

\subsection{Dressed Polyakov loop}

It is a much more non-trivial task to find an order parameter for confinement.
The desirable object would be the Polyakov loop which can be connected to the
free energy of a static quark, and is thus a description for confinement.
Since we can not directly access the Polyakov loop from solutions of DSEs
we will use the so-called dual condensates. They are defined as

\begin{equation} \label{eq:dualCondensate}
\Sigma_n = \int_0^{2\pi}\frac{d\varphi}{2\pi}e^{-i\varphi n} \langle\bar\psi\psi\rangle_\varphi,
\end{equation}
where $\langle\bar{\psi}\psi\rangle_\varphi$ 
is the quark condensate evaluated at generalized $U(1)$-valued boundary conditions 
$\psi(\vec{x},1/T)=e^{i\varphi}\psi(\vec{x},0)$ with $\varphi \in [0,2\pi[$.
In \cite{Gattringer:2006ci,Synatschke:2007bz,Bilgici:2008qy}
it has been shown that $\Sigma_{\pm 1}$ contains the ordinary Polyakov loop
as well as contributions from spatial detours. It has therefore been called
the dressed Polyakov loop. Most notably it is sensible to center symmetry,
just like the conventional Polyakov loop.
It is therefore an order parameter for confinement that is accessible by
continuum functional methods, since only the dressed quark propagator is necessary
as an input \cite{Fischer:2009wc}.

With Eqs.~(\ref{eq:condensate},\ref{eq:dualCondensate}) we have identified
order parameters that rely only on the quark propagator.
We will now define the truncation scheme that we will use to get our hands on
not only the quark but also the gluon propagator.

\section{Truncation scheme}

\begin{figure}
\centering
\includegraphics{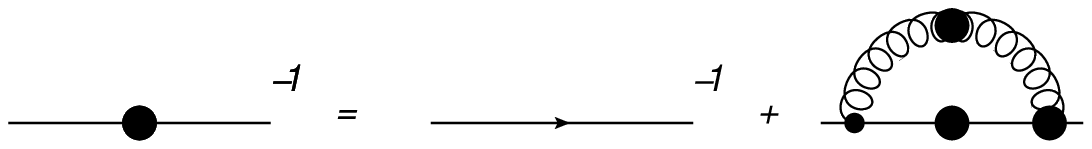}
\includegraphics{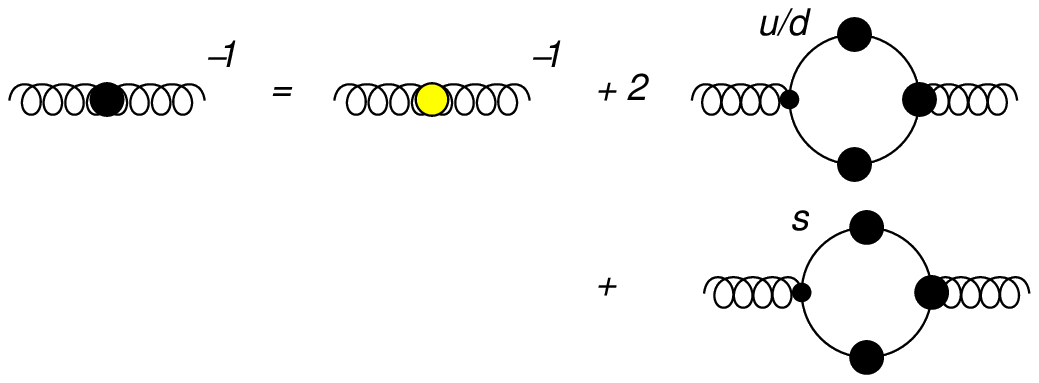}
\caption{The coupled Dyson-Schwinger equations for the quark and gluon propagators.}
\label{fig:DSEsQuarkGluon}
\end{figure}

The upper diagram in Fig.~(\ref{fig:DSEsQuarkGluon}) shows the quark DSE.
This equation defines the fully dressed quark propagator.
On the right hand side we have the self-energy which depends on the fully dressed
gluon and quark-gluon vertex. We need a prescription for these objects in order
to get a closed set of equations, this will be given in the following.
The challenge of in-medium Dyson-Schwinger equations is to include the relevant medium
effects in the applied truncation scheme.
We will here use the same construction for the quark-gluon vertex as in
\cite{Fischer:2009wc,Fischer:2009gk,Fischer:2011mz,Fischer:2012vc}:

\begin{eqnarray}
\Gamma_\mu&=&\gamma_\mu\cdot\Gamma(p^2,k^2,q^2) \cdot 
\left(\delta_{\mu,4}\frac{C(p)+C(k)}{2} + \delta_{\mu,i}\frac{A(p)+A(k)}{2} \right),\\
\Gamma(p^2,k^2,q^2) &=& \frac{d_1}{d_2+q^2} \!
 + \!\frac{q^2}{\Lambda^2+q^2}
\left(\frac{\beta_0 \alpha(\mu)\ln[q^2/\Lambda^2+1]}{4\pi}\right)^{2\delta},
\end{eqnarray}
here $q$ is the gluon momentum while $p$ and $k$ are the quark momenta.
This {\it ansatz} contains the first part of the Ball-Chiu construction and is thus dependent on the medium.
On the other hand, the dressing function $\Gamma$ is constant in $T$ and $\mu$.
This works well up to $T_c$ but leads to an
over-estimation of chiral symmetry breaking in the quark-gluon plasma as will be shown below.

The parameters are fixed to reproduce the pion decay constant in the vacuum and turn
out to be $d_1=7.5$ GeV$^2$, $d_2=0.5$ GeV$^2$ and $\Lambda=1.4$ GeV.
In the UV we have $\delta=-9\frac{Nc}{44N_c - 8N_f}$ and $\beta_0=\frac{11N_c-2N_f}{3}$.

For the gluon propagator our strategy is to start with quenched QCD. Here, lattice
calculations are up to now the most reliable source for the Landau gauge and fully
temperature dependent propagator \cite{Fischer:2010fx}. We have used this input to calculate
the first (second) order phase transition in $SU(3)$ ($SU(2)$) gauge theory in \cite{Lucker:2011dq}.
To be able to correctly describe the crossover in unquenched QCD and to go
to finite chemical potential, we add the quark loop from the gluon DSE to the
quenched propagator.
In Fig.~(\ref{fig:DSEsQuarkGluon}) the resulting set of equations that we have solved is shown
diagrammatically, where the first term on the right hand side of the gluon DSE denotes the quenched propagator.
A strange-quark loop is included for the $N_f=2+1$ case.
This procedure for unquenching the gluon propagator neglects all quark loops inside the Yang-Mills self-energies.
In the vacuum this leads to an error on the 5 percent level.

The quark loop at finite temperature can be split into longitudinal (L)
and transversal (T) parts:

\begin{equation}
\Pi_{\mu\nu}(p) = P^T_{\mu\nu} \Pi^T(p) + P^L_{\mu\nu} \Pi^L(p),
\end{equation}
where one notes that $\lim_{p\rightarrow 0}p^2\Pi^T(p)=0$ but

\begin{equation}
\lim_{p\rightarrow 0}p^2\Pi^L(p) = \frac{m_{Th.}^2}{2}, \label{eq:thMass}
\end{equation}
i.e. the longitudinal part produces a thermal (or Debye) screening mass.
In the color superconducting phase the transversal part would generate
a Meissner mass in a similar way.

\section{Results}

\begin{figure}
\includegraphics[width=.48\textwidth]{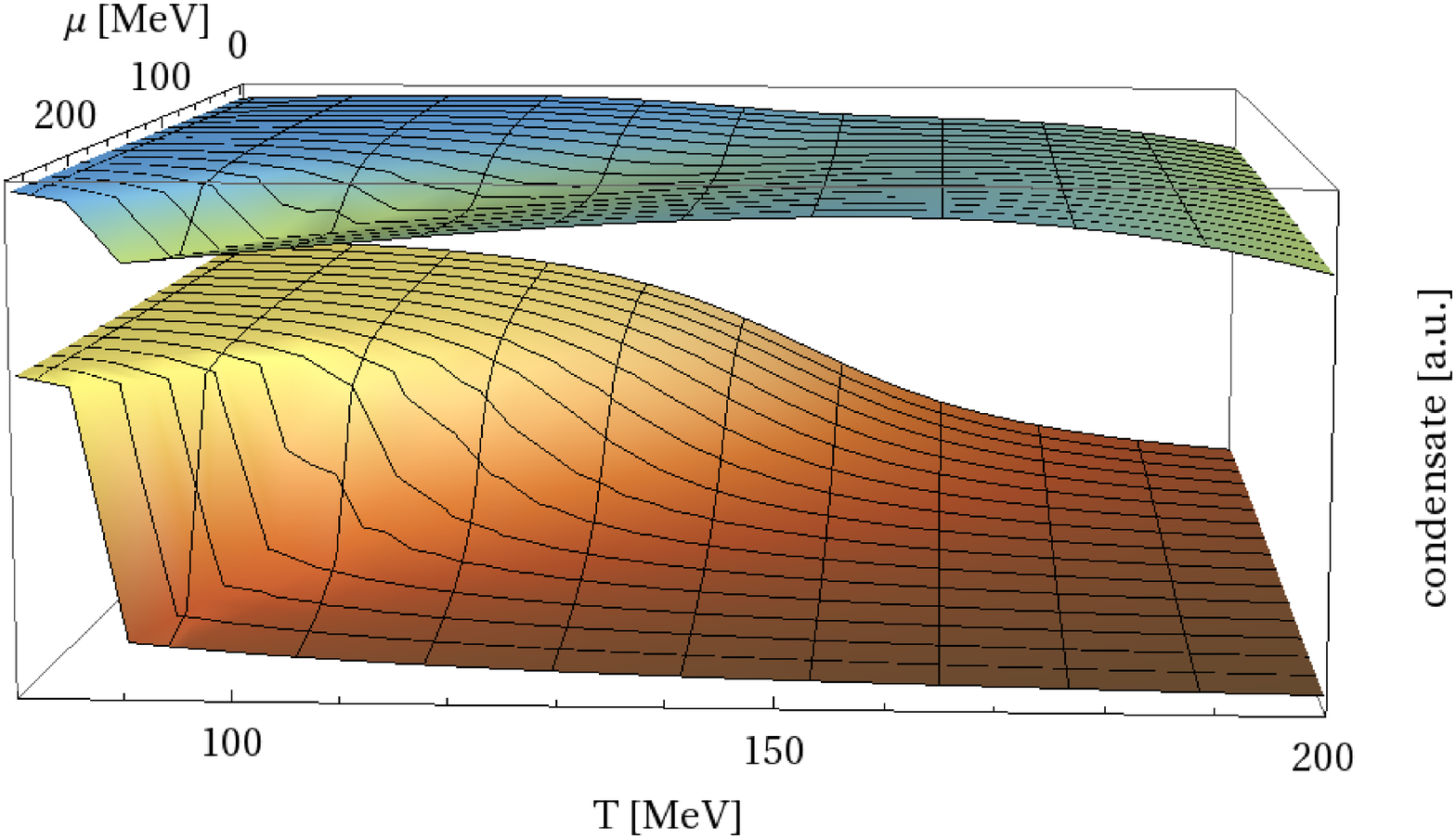}
\hfill
\includegraphics[width=.48\textwidth]{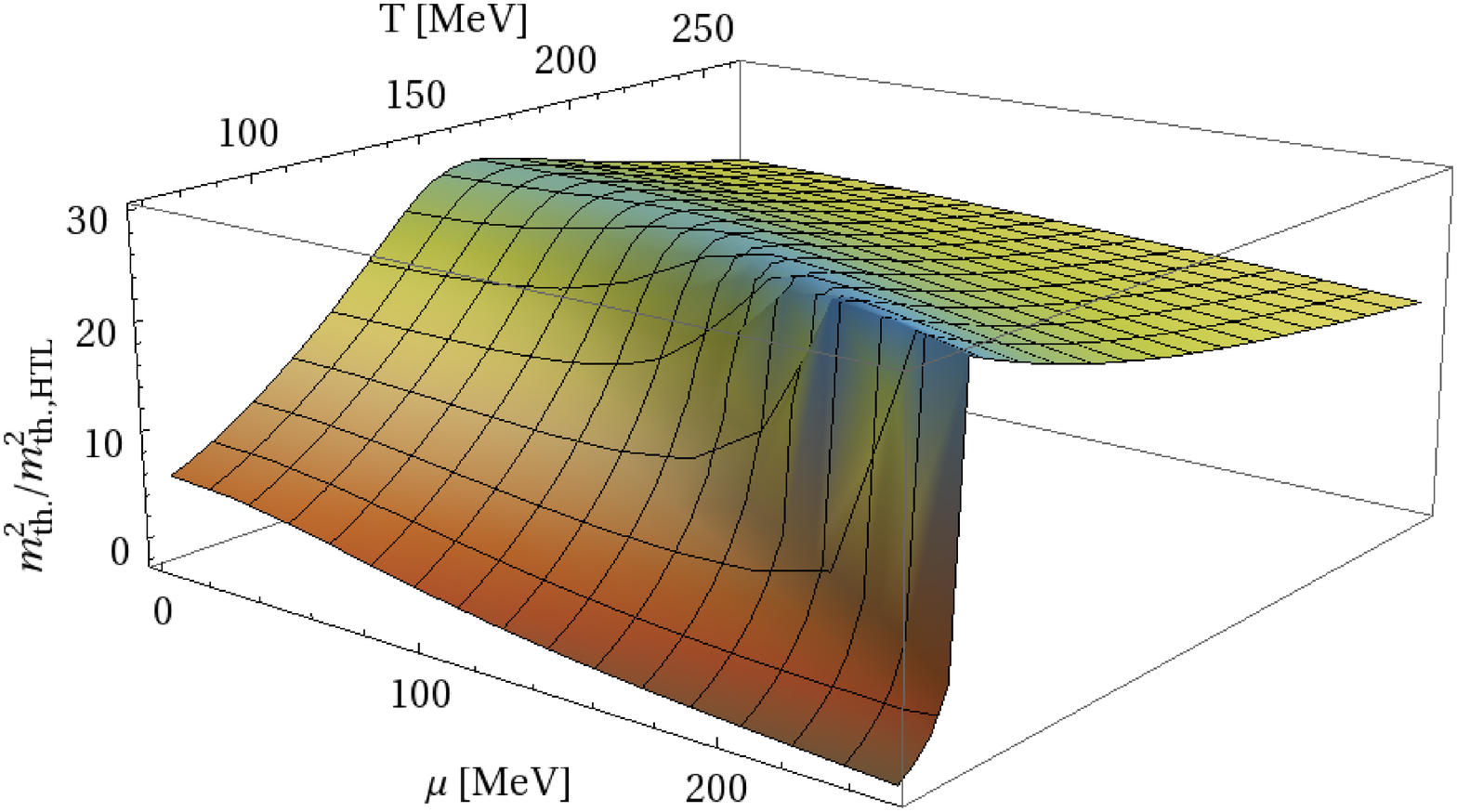}
\caption{The left figure shows the light and strange quark condensates as the lower
and upper surfaces, respectively.
The right figure shows the thermal mass contribution from the quark loop,
normalized by $T^2 + \frac{\pi}{3}\mu^2$, the asymptotic value.
Both figures are adopted from Ref.~\cite{Fischer:2012vc}.}
\label{fig:condsAndMmth}
\end{figure}

We present results from a calculation of the quark condensates and the thermal mass for
$N_f=2+1$ QCD in the left side of Fig.~(\ref{fig:condsAndMmth}).

What we find from the light quark condensate is a crossover for small chemical potential, which becomes
stronger with rising $\mu$ and eventually turns into a first order phase transition.
The strange quark condensate reflects this change due to the coupling to the light quark via the gluon
propagator. However, it continues to decrease at larger temperatures due to the larger strange
quark mass.

The contribution to the thermal gluon mass due to the quark loop, Eq.~(\ref{eq:thMass}),
is shown to the right in Fig.~(\ref{fig:condsAndMmth}), normalized by its asymptotic HTL/HDL behaviour.
It is small in the hadronic phase, where it is suppressed by the inverse quark mass.
Around the phase transition it rises strongly and approaches the asymptotic value
above $T_c$.
For large $\mu$, at the first order phase transition, we observe a jump, which is inherited from the
jump in the quark propagator.
The quark loop reduces the gluon strength, and therefore also the quark self-energy.
A smaller self energy in turn leads to a larger quark loop, and therefore
the back-coupling of quark and gluon has an accelerating effect on the phase transition.
This acceleration will be important to reproduce lattice data, which we do in the following.

\begin{figure}
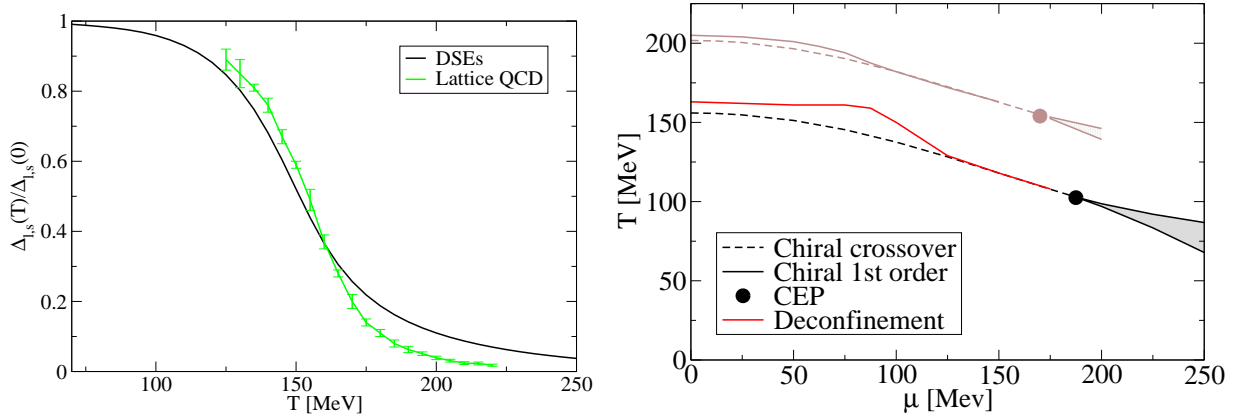

\centering
\includegraphics[width=.48\textwidth]{compareLattice}
\hfill
\includegraphics[width=.5\textwidth]{phaseDiagNf2p1}
\caption{In the left figure we compare the regulated and normalized quark condensate to corresponding
results from lattice QCD \cite{{Borsanyi:2010bp}}.
In the right figure we show the phase diagram for $N_f=2+1$ QCD. In lighter colors we also show the
results for $N_f=2$ QCD (upper lines).
Both figures are adopted from Ref.~\cite{Fischer:2012vc}.}
\label{fig:compLattice}
\label{fig:phaseDiagram}
\end{figure}

From the light and strange quark condensates we can obtain the quantity

\begin{equation}
\Delta_{l,s} = \langle\bar\psi\psi\rangle_l - \frac{m_l}{m_s}\langle\bar\psi\psi\rangle_s,
\end{equation}
which is finite since the divergent term $m\Lambda^2$ in $\langle\bar\psi\psi\rangle$ cancels.
It can therefore be used in comparisons with different approaches like lattice QCD.
This comparison is shown in the left part of Fig.~(\ref{fig:compLattice}).
It shows a rather good agreement up to $T_c$ and an overestimation of the remaining
chiral symmetry breaking in our approach above $T_c$.
This we attribute to the constant vertex strength,
which should be temperature dependent in a refined model.
Nonetheless we find a critical temperature of $T_c \approx 156$ MeV in good agreement with
the lattice \cite{{Borsanyi:2010bp}}, and a similar steepness of the crossover.

Finally we show the main result of Ref.~\cite{Fischer:2012vc}, the phase diagram of $N_f=2$
and $N_f=2+1$ QCD in the right part of Fig.~(\ref{fig:phaseDiagram}).
We find a critical end-point at $(\mu,T)\approx(190$ MeV$,100$ MeV$)$ for $N_f=2+1$, which
is certainly outside the area that can be accessed by extrapolations from lattice data.
The (pseudo-)critical temperature for deconfinement lies slightly above that
of chiral restoration in the crossover regime, and coincides close to and at the CEP.
Fig.~(\ref{fig:phaseDiagram}) also shows in lighter colors the phase diagram for
the two flavor case, where $T_c \approx 200$ MeV. The impact of the strange quark
is evidently a reduction of $T_c$ by about $50$ MeV while the CEP moves only slightly
to larger $\mu$.

\section{Summary}

We discussed a truncation scheme to the Dyson-Schwinger equations of QCD
that explicitly couples the quark and gluon
propagators. This makes the gluon sensitive to the chiral phase transition
via the thermal mass contribution from the quark loop.
In this approximation we showed that the condensate is in reasonable
agreement with lattice simulations.
For the phase diagram we find a CEP at $\mu/T \approx 1.9$, and
nearby chiral and deconfinement phase transitions.

\ack
We thank the organisers of the FAIRNESS workshop 2012.
This work has been 
supported by the Helmholtz Young Investigator Grant VH-NG-332 and the Helmholtz 
International Center for FAIR within the LOEWE program of the State of Hesse.

\end{document}